\documentclass [preprint,aps]{revtex4}
\usepackage{bm}
\begin{document}
\title{Study of Coherent and Incoherent Plasma Emissions: Role of Plasma Wave Excitation and Breaking}   
\author{A. S. Sandhu, G. R. Kumar}
\address{Tata Institute of Fundamental Research, 1 Homi Bhabha Road, Mumbai 400 005, India,}
\author{S. Sengupta, A. Das, and P. K. Kaw}
\address{Institute for Plasma Research, Bhat, Gandhinagar, Ahmedabad 382428, India}
\date{\today}

\begin{abstract}
	Here we experimentally  map the dynamics of electron plasma waves in laser solid interaction. We do time resolved measurements of second harmonic and hard X-ray generation from interaction of intense ($10^{16} W cm^{-2}$, 100 fs, 800nm) laser with a pre-plasma generated on a solid surface. The parameter space explored in this time resolved study includes variation of scale length, laser polarization and laser intensity in conjunction. These measurements done together  brings novel features of strongly driven electron plasma wave behavior, which have not been explored experimentally so far. We model the results in terms of Resonance Absorption and Wave-Breaking mechanisms. The Harmonic and X-ray emission show contrasting behaviour, which indicates pitfalls in trying to increase harmonic efficiencies by brute force. However by simple adjustments,  we observe that hard X-rays can be enhanced or controlled upto two orders of magnitude and second harmonic upto one order of magnitude under optimum conditions. These results should help us understand the governing mechanisms for short wavelength generation and fast particle generation to develop more efficient sources for application purposes.	
\end{abstract}


\maketitle

The advent of short pulse and high peak power laser technology has fuelled development of new sources of light and fast particles. This also has resulted in development of new ideas and physics behind interaction of high intensity light with matter. There is immense interest in understanding and increasing efficiency of High Harmonic Generation (HHG) from different media like gases, solids, clusters and liquid drops etc. HHG is the most promising way of creating XUV \cite{Kapteyn,Solid,Others} and water-window photon sources \cite{Krausz} for applications in lithography, real time imaging of biological samples. Similarly the electron and ion acceleration  using high power lasers has resulted in advances in realizing Tabletop particle accelerator schemes \cite{Hegelich}. Another important area  concerns the yeilds and applications of short pulse X-ray sources using laser-solid interaction \cite{Bastiani},\cite{Linde} for probing atomic and nuclear motions. 

In intense laser-solid iteraction, which is our main concern in this work (although the reults may be valid for liquids, clusters as well), it is a common knowledge that the collective processes like resonance absorption and plasma oscillations play an important role in high harmonic and X-ray generation. The resonance absorption phenomenon has been expensively studied in literature. Since it is leads to strong excitation of electron plasma waves (EPW) near critical density, the growth and decay of these plasma waves thus plays crucial role. Such dynamical aspects of plasma waves have been theoretically been studied extensively using analytical and numerical techniques, although experimental verification and understanding is still lagging much behind. As an example the concept of electron plasma wavebreaking invoked first by Dawson \cite{Dawson} is particularly intresting and significant \cite{Coffey, Mulser} as it has been understood as the mechanism responsible for the generation of fast particles observed in laser-solid experiments. However hardly any direct experimental evidence of the effect in context of high intensity laser plasma interaction. There has been one experimental report \cite{Gizzi} invoking wavebreaking (WB), however, it does not go into details of the phenomenon, but stops merely at hinting towards the existence of wavebreaking phenomenon. In view of ever increasing intensities of lasers it seems inevitable that plasma waves of large amplitudes do not break. Thus undoubtedly, there is immense importance of EPW dynamics in deciding the efficiencies of plasma emmission both of coherent type (High Harmonic Genration) and incoherent type (fast particle and hard X-ray generation). 

We here do an detailed experimental study and modelling of the EPW phenomenon over large parameter space. The parameters that were simultaneously varied in conjunction for this time resolved study included the scale length of plasma, the laser polarization, the pump laser internsity, the prepulse laser intensity. The observables include pump reflectivity, probe reflectivity, Second Harmonic Generated (SHG) by the pump, Hard-Xray emmitted and the spectrum of the reflected fundamental and SHG. We establish important mechanisms, which operate and control harmonic and X-ray emissions from plasmas. The simple optimization of conditions is shown to lead to X-ray enhancements by two orders of magnitude and second harmonic by one order of magntitude. The results also show that parameter space for optimisation of X-ray and harmonic yields can be very different and the difference results from process of plasma wave breaking. The results also indicate pitfalls in simple minded approach to increase harmonic efficiencies by brute force increase of laser intensity.

 	The experimental set-up consists $10^{16} W cm^{-2}$ pump (100 fs, 800nm) incident at $45^{\circ}$ with respect to target normal. The prepulse is 50 times weaker than pump and at near normal incidence. We monitor the input laser, reflected pump, reflected probe with photodiodes (Hamamatsu) and the second harmonic generated by pump is monitored with PMT (Hamamatsu). The SHG emmission was observed to be specular in our experiments. The target used in this study is solid glass (BK7) $2" \times 2" \times 5mm$ pieces. Finally NaI (Tl) detector appropriately  gated with respect to laser, placed at $22.5^{\circ}$ with respect to target normal is used to measure the hard X-ray emissions from the plasma in the energy range from 30 to 500 keV. This experiment involved variation of scale length of preplasma and polarization plane of main laser pulse and intensities of pump and probe, unlike any previous studies where at most one variation was considered at a time leading to incomplete picture.

First we compare the hard X-ray and SHG measurements as a function of prepulse to main pulse time delay (i.e. as function of scale length of preplasma) at fixed polarization of pump (p-polarization), fixed pump intensity ($ 10^{16} W cm^{-2}$), fixed probe intensity (50 times weaker than pump).  The results are shown in figure 1. We have plotted integrated hard X-ray yeild over 30-500keV and second harmonic efficiency which defined as $\eta = I(2\omega)/I^{2}(\omega)$ with time delay. 
In both the plots the data are normalized with respect to yeilds obtained with the pump alone acting, thus the baseline value for pump alone acting is normailzed to unity in each case. The most conspicuous feature is contrary behaviour of Second Harmonic generation (SHG) and X-rays near vicinity of 24 ps delay. The X-ray yield is seen to enhance by a factor of 140 ,i.e. more than tw order of magntitude, when compared to yield with pump alone indicating large amount of fast electron generation near 24 ps. The SHG on the other hand shows a dip around 24 ps thus exhibiting perfect anti-correlation with X-ray data. The maximum enhancement in SHG at the peak values around the dip is 3 times when normalized with respect to main pulse acting alone. At the dip the SHG is still about 1.5 times the value without prepulse. The baseline, which has been normalized to unity here, corresponds  to absolute SHG efficiency of $ 5 \times 10^{-6}$ with p-pump and at least 50 times lower for s-pump alone.

Next instead of comparing total yeilds of X-rays with and without prepulse we now spectrally resolve the X-ray yeilds and compare hot electron temperatures at different positions. One with only pump acting which corresponds to points on baseline of figure 1(b) and other with prepulse dealy of 24 ps that at position where X-ray yeild enhances by a factor of 140.
The result shown in figure two shows enhanced hot electron temperatures at 24 ps. Comparison of the hot electron spectra shows that; in absense of prepulse, we get 3 keV as main temperature component, while the second component is almost insignificant as seen from statistical weights. With the prepulse at 24 ps, the we require two temperatures for good fit, 5 and 37 keV, the second temperature being very prominent unlike the first case. This indicates strong RA is operational and large quantity of fast particles is being generated, with optimised prepulse delay of 24 ps. A resonable explanation of these results lies in the phenomenon of excitation of plasma waves via Resonance Absorption (RA). RA is crucially dependent on the scale length of the plasma and the component of electric field parallel to density gradient driving the plasma waves i.e. polarization of the pump pulse. It is expected that at certain optimum length scale there will be maximal RA, where plasma waves are driven very hard. Similarly in going from S- to P-polarization we are effectively increasing the E field component along density gradient, which is driving plasma waves under conditions of resonance absorption. 

We next establish these facts quantitavely using the reflectivity of the main pulse to deduce the scale lengths of preplasma(Fig 3). The top x-axis is time delay and bottom axis represents scale length resulting from prepulse deduced by numerically calculated reflectivity (solid curve) solving  Helmoltz equation in a sharp density profile \cite{Milchberg}, with a collision frequency of $0.05\omega$. Thus the scale length at 25 ps where we see phenomenon of enhancement of X-rays and simultaneous dip of SHG is $\sim \lambda$ under our conditions. This region around $L \sim \lambda$  (time delay 15-25 ps) corresponds to maximum absorption and hence large number of plasma waves are strongly driven in this resonance absorption region. (Remember the defination of scale length is important, here we are deducing actually total plasma length, scale length may be smaller than this, depending on the profile used.)

Now that resonance absorption(RA) is clearly established, we attempt to explain the figure 1 and 2. As noted earlier in near resonance region we do expect large amplitude plasma waves driven strongly by the laser. Hence these plasma waves when damping can be invoked to explain enhanced X-ray yeilds and temperatures. However this explanation is clearly not sufficient when to attempt to exlplain the SHG data, in particular the strong dip around eactly the position where large amount of fast particles are being generated. Thus we have to look more carefully at the behavior of strongly driven plasma waves. As the SHG current is assumed to come from two contribitions one term proportional to pondemotive force which will be effective both for s and p-polarization and other term proportional to  $ \nabla . n_{e} E$ which is strongly active for plasma waves driven by p-polarized laser.  Now if the plasma waves start breaking when driven too hard the second term which is main contribution will be disrupted and hence we will get decrease of SHG. Since we are actually driving plasma very very hard hence it is not unexpected that wave breaking will set in when density fluctuation becomes of the order of unperturbed density itself. Thus the wavebreaking phenomenon which is theoretically understood very well has to be invoked here to undertand the results. In addition here we can measure the exact conditions for the onset of wave breaking. This should help in testing the theoretical and numerical wave breaking models .

Now we model the physics of wavebreaking relevant to our conditions and parameters. The starting point is the equation 
$ k_{p} eE_{p} /m \omega^{2} \ge 1 $ for breaking of the plasma wave i.e. the ratio of the amplitude of the oscillation of the electron to the wavelength of the plasma wave, where $k_{p}, E_{p}$ is the wave vector and electric field of the plasma wave.  Further observing $E_{p} = E_{d}/\epsilon$, where $E_{d}$ is the driving electric field and $\epsilon$ is dielectric constant. The dielectric constant will have a non zero minimum value at critical layer, (due to finite width of resonance region?) and minimum value is obtained by averaging over half a wavelength aroung resonance. We get $\epsilon_{min} = \pi/(2k_{min}L) = \lambda_{p}/4 L$, where L is plasma scale length. Now we use $k_{p} = 2\pi/\lambda_{p}$, and $\lambda_{p} = \lambda_{de}^{2/3} L^{1/3}$ for thermal plasma waves. Now putting this together and using the defination $E_{th} = k T_{e}/e \lambda_{de}$ substituting in the original condition, we get the wave breaking condition (WBC) as $(E_{d}/E_{th})(L/\lambda_{de})^{1/3} \ge 1 $. The driver field can either be calucated numerically again using solution of Helmoltz calculation in steep density profile done above for fitting pump reflectivity or one can simply get the resonance absorption from denisov function. Then assuming a temperature of 200 eV, we plot the WBC for different peak laser intensities in figure 4. The flat line at $WBC = 1$ represents the critical limit necessary for wavebreaking. The denisov function relfectivity is also plotted  at the bottom. The WBC for low intensities is not satisfied and hence waves will not break for intensity of order $10^{14} W cm{-2}$. However as we increase the intensity the WBC is satifies over larger and larger scale lengths. At the maximum intensity $10^{16} W cm{-2}$ waves can break over almost upto $1.5 \lambda$. Hence we verify that wavebreaking does occur comfortably over our range of parameters. In addition this illustrates that wavebreaking is ubiquitous in intense laser plasma interaction studies done around or above $10^{16} W cm{-2}$.

Now we revisit figure 1 with these results. The hard Xray enhancement is explained as during wavebreaking the energy lost by the wave will be converted to kinetic energy of electrons, thus the accelerated electrons will result in enhanced bremsstrahlung, in agreement with our experimental results. In case of SHG we can understand that as the scale length changes, we pass through resonance absorption region where plasma wave amplitude increases to high value leading to breaking of plasma waves. As we know wave breaking results in disruption of the non-linear source current thus explaining the dip observed in SHG. However still we need to explain the non zero value of 1.5 SHG efficiency at dip position at 24 ps. The reason for related to the fact for a given laser pulse which typically has 40-50 laser cycles, each one exciting plasma waves. Thus at each cycle we can define the intensity of inident wave and hence driver field for the plasma wave. Thus we have to view it as highly dynamical situation in 100 fs duration. As we increase peak intensity of laser pulse; at certain critical intesity the plasma waves related to peak of the laser pulse start to break thus reducing SHG, however, the wings of laser still produce SHG efficiently. As pulse peak value of intensity is increased further more and more laser cycles near peak contribute to plasma wavebreaking. However still there is some non-broken waves at the wings which will keep SHG to non zero value. If we make simple estimate in case of peak pulse intensity of $10^{16} W cm^{-2}$, the laser cycles with peak intensity below 1/50 of peak will not break as seen from the figure 4. Now if we estimate for a gaussian pulse roughly 1/10 of energy of pulse is contained in the wings below the wave breaking threshold which will continue to yeild SHG. This clearly is not enough as the dip at 24 ps is only factor of two lower that the maximum SHG efficiency observed, where energy in wings is expected to yeild much lower SHG efficiency. The explanation behind this lies in the fact that even in the case of maximum SHG position in figure 1, there is enough wavebreaking present, which keeps the maximum value of SHG signal low. Hence one should compare the efficiency for 24 ps dip position to the maximum value which would have been there in totally wave breaking free experiment.

To make the situation clear and to substantiate above claims we now study the behavior of SHG with scale length at low and high intensity cases. Figure 5 shows three cases: p-polarized pump at high intensity ($10^{16} W cm^{-2}$), p-polarized pump at low intensity ($2 x 10^{15} Wcm^{-2}$) and s-polarized pump at high intensity ($10^{16} W cm^{-2}$)
For the sake of comparison we have multiplied the efficiencies in each case by factor indicated in the figure 5. This is done in such a way that value for pump alone acting is each case is same as for highest intensity p-polarized pump case which as shown already in figure 1 to be normalized to unity. Thus the curve for s-polarized case in reality is a lower in SHG efficiency value by a factor of 65  that of high intensity p-polarized case and the low intensity p-polarized case is a factor of 20 below that of high intensity case. It is obvious from the figure 5 that at low intensities 
Indeed if we reduce the peak laser pulse intensity we see that maximum SHG efficiency in low intensity case  is as large as a factor 13 compared to pump alone acting. Thus this experiment is relatively wavebreaking free version of high intensity strongly wavebreaking case, where the maximum SHG efficiency goes only upto factor of 3 before dipping. Also now the finite non zero SHG efficiency value of 1.5 at the dip at 24 ps in high intensity plot, when compared to maximum value of 13 in low intensity relatively wavebreaking free yeilds a contrast of about a factor of 1.5 to 13, which can be explained to be due SHG generated in the wings of the pulse. This disscussion on comparison of high and low intensity cases also make it clear that wavebreaking is taking place not only in the scale length regime where SHG shows a dip, but over a large range of scalelengths. Further the scale length region of validity of WBC increases with increasing intensity.
Finally SHG efficiency for s-polarization although very low can be explained to be due to pondermotive effect and this effect can also keep SHG to a finite value. In addition source of SHG in case of s-polarization is due to imperfect polarization state of the laser (purity 99 percent) and focussing geometry effects over the spatial spot size of the laser.

Further we now see the intensity scaling of SHG at fixed delay positions (5ps,21 ps) and polarization state (p-pol). The results are plotted in figure 6. Also plotted are the fits assuming SHG yeild varies as square of incident intensity. We observe that the scling deviates from exponent value of 2, at intensity equivalent 0.1 of the maximum  intensity. The deviation from square scaling is also observed to be earlier in case when the delay is at 21 ps i.e. the dip position and final difference in values of fit and observed values is also much larger, than observed at 5 ps.

Now same arguements used above for scale length variation study of wavebreaking can be used in case of polarization variation. As we change the polarization from S to P we are increasing the driving field for elctron plasma waves and hence their amplitude. This ultimately leads to breaking of the waves when a critical limit is reached.
Figure 7 shows the dependence of X-ray yield as a function of pump polarization for two cases, with and without prepulse. With prepulse at 24 ps (maximum yeild position in figure 1), we observe that going from S-pol to P-pol we can enhance X-ray yield by two orders of magnitude. Without prepulse however maximum change is atmost a factor of 2. Thus in conjunction with a prepulse, laser polarization can be a very effective and simple way to control X-ray yield over a wide dynamic range. Inset of fig 7 shows the ratio of two yields above i.e. with and without prepulse as function of polarization. The enhancement is maximum for P-polarization and insignificant for S-polarization. 
This demonstrates of a simple way to control X-ray yeilds over two orders of magntitude. Merely by rotating the half wave plate one can tune X-ray yeilds to a factor of 100 if there is a prepulse present at right delay. If however prepulse were not present the rotation of half wave plate will not yeild even a factor 2 variation.

Same effect can be observed in SHG also where the wavebreaking is brought about clearly. We compare  polarization dependence of SHG ahown in figure 8. The SHG behaviour without and with a prepulse at 7ps, which is position of peak enhencement of SHG is completely opposite to that of X-rays. There is enhancement in both cases i.e. with and without prepulse for all polarization states, however the relative enhancement (the ratio of yields with and without prepulse) is maximum for S-polarization ($\sim 14$) whereas only 3 for P-polarization (inset fig 3). Comparing insets of figure 7 and figure 8, we have an important observation namely that the enhancement ratio curve as a function of polarization is opposite for X-rays and SHG due to wavebreaking effects.

Finally the spectrum of SHG in direction collinear with reflected pump is shown in figure 9. At negative delay the spectrum is smooth. For p-polarization at positive delays as we go near the dip position the spectrum starts becoming more and more distorted and split.

In conclusion, the novel time resolved simultaneous study of SHG and X-rays as a function of scale length and laser polarization elucidates features of electron plasma wave dynamics. The Harmonic and X-ray emission yeilds are observe over wide range of parameters. We interpret the results in terms of Resonance Absorption and Wave-Breaking mechanisms. The modelling of results is in good agreement with observed results. We propose SHG is a very sensitive indicator of wave breaking and can be useful tool for understanding plasma wave dynamics. Under optimum conditions we observe that hard X-rays can be enhanced by a factor of 100 and second harmonic by factor of 10. The plasma scale length and laser polarization when used in conjunction, provide a wide dynamic range and are thus most effective as controlling parameters of short wavelength sources for application purposes. 

We would like to thank D. Mathur, M. Krishnamurthy, P. P. Rajeev, A. K. Dharmadhikari, J.A. Dharmadhikari and M. Anand for discussions and useful suggestions. We also thank S. Narayanan, S. Bagchi and Aoife Bharucha for help.

\newpage

\begin{figure}
\caption{(a) The efficiency of SHG as a function of time delay, (b) Integrated yield of X-ray photons from 30 keV to 500 keV, both normalized with respect to yield in absence of prepulse i.e. when pump is acting alone.}
\end{figure}

\begin{figure}
\caption{Hot electron temperature fits (a) without prepulse are 3 keV and 20 keV. The second component is almost insignificant (b) Hot electron temperature in case of prepulse at 24 ps, are 5 and 37 keV. The second temperature is prominent and has much larger statistical weight than in first case.}
\end{figure}

\begin{figure}
\caption{The reflectivity of the pump pulse with delay (scale length) of the preplasma. The solid line: numerically calculated absorption.}
\end{figure}

\begin{figure}
\caption{The wavebreaking condition (WBC) parameter derived for different laser intensities. Also plotted is the the reflected caluculated by assuming linear profile (Denisov Formula).}
\end{figure}

\begin{figure}
\caption{Normalized behavior of SHG efficiency as a function of time delay for s - pump and p- pump at low and high intensities. For the sake of comparison we have multiplied the efficiencies in each case by factor indicated; s-polarized case in reality is a lower in SHG efficiency value by a factor of 65 and the low intensity p-polarized case is a factor of 20 below that of high intensity case shown in the figure.}
\end{figure}

\begin{figure}
\caption{The scaling of SHG yeild with intensity over 2 orders of magnitude in intensity. Clearly the behavior is exactly intensity squared at low intensities, at high intensities it deviates. For time delay 21 ps the deivation starts are lower intensity than in case of 5 ps.}
\end{figure}

\begin{figure}
\caption{Polarization dependence of X-ray yield with and without prepulse; enhancement 100 and 2 respectively (normalized s.t. yield in absence of prepulse at S-pol is unity). Inset: Relative Enhancement i.e Ratio of yield with and without prepulse}
\end{figure}

\begin{figure}
\caption{SHG efficiency as function of polarization, with and without prepulse (normalized s.t. SHG efficiency in absense of prepulse at S-pol position is unity). Inset: The relative enhancement i.e ratio of yeild with prepulse to without prepulse,  maximizes for S-polarization unlike X-rays.}
\end{figure}

\begin{figure}
\caption{The spectrum of second harmonic collinear with reflected pump. We notice that s -polarization yeilds smooth spectrum for SHG at all delays. However for p-polarization the spectrum is smooth intially and gets progressively more distorted and split as we increase the delay.}
\end{figure}


\newpage 

 
\begin{references}

\bibitem{Kapteyn} A. Paul {\it et al.}, Nature $\bf{51}$, 421 (2003) , R. A. Bartels {\it et. al.}, Science $\bf{297}$, 376 (2002).

\bibitem{Solid} M. Zepf {\it et. al.}, Phys. Rev. E $\bf{58}$, R5253 (1998), A. Tarasevitch, Von-der-Linde, {\it et. al.}, Phys. Rev. A $\bf{62}$, 023816 (2000).

\bibitem{Others} Vrakking, Ditmire etc, {\it et. al.}, PRL $\bf{xx}$, xxx (xxxx).

\bibitem{Krausz} Ch. Spielmann {\it et al.} Science $\bf{278}$, 661 (1997).

\bibitem{Hegelich} M. Hegelich et. al. Phys. Rev. Lett. 89, 085002 (2002), J. D. Kmetec et. al. Phys. Rev. Lett.  68, 1527 (1992),V. Kumarappan et. al. Phys. Rev. Lett 87, 085005 (2001).

\bibitem{Bastiani} S. Bastiani, {\it et al.}, Phys. Rev. E $\bf{56}$ 7179 (1997) 

\bibitem{Linde} K. S-Tinten {\it et al.}, Nature $\bf{422}$, 287 (2003).

\bibitem{Dawson} J. M. Dawson Phys. Rev. $\bf{113}$, 383(1959).

\bibitem{Coffey} T. P. Coffey, Phys. Fluids $\bf{14}$,1402 (1971).

\bibitem{Mulser} A. Bergmann and P. Mulser, Phys. Rev. E $\bf{47}$, 3585 (1993).

\bibitem{Gizzi} L. A. Gizzi, D. Guiletti and A. Guiletti, Phys. Rev. Lett. $\bf{76}$ 2278 (1996).

\bibitem{Milchberg} H. M. Milchberg and R. R. Freeman, J. Opt. Soc. Am. B $\bf{6}$ 1351 (1989).

\end{references}
\end{document}